\documentclass[seceq,preprint]{ptptex}

\usepackage{graphicx}

\newcommand{\nn}{\nonumber}
\newcommand{\p}{\partial}
\newcommand{\inv}[1]{\frac{1}{#1}}
\newcommand{\f}[2]{\frac{#1}{#2}}
\newcommand{\R}{{\mathbf{R}}}
\newcommand{\Z}{{\mathbf{Z}}}
\newcommand{\C}{{\mathbf{C}}}
\newcommand{\abs}[1]{|#1|}
\newcommand{\dis}{\displaystyle}

\renewcommand{\Im}{{\rm Im}}

\preprintnumber[3cm]{
KUNS-1805\\ hep-th/0210039}

\title{
Asymptotic Behavior of Gravitational Wave on the Brane World%
}

\author{
Yutaka \textsc{Murakami}\footnote{E-mail:
  murakami@gauge.scphys.kyoto-u.ac.jp}%
}

\inst{
Department of Physics, Kyoto University, Kyoto 606-8502, Japan
}

\abst{
We consider the Cauchy problem of the gravitational wave with the
initial data distributed only around the brane in the one brane model
of Randall and Sundrum, and examine its behavior as
$t\rightarrow\infty$ .  Then we find its leading behavior is $t^{-6}$
unlike an ordinary flat 5-dimensional space-time.  Such a signal shows
that the Huygens principle is violated
on the 4-dimensional brane world even asymptotically and also shows
the difference between compact and non-compact brane worlds.  Some
comments are also given related to AdS/CFT correspondence .
}

\begin{document}

\maketitle

\section{Introduction}
\label{sec:intro}

Recently, inspired by the string theory,\cite{HW} \  there are many
works called as brane world scenario,\cite{ADD,AADD,RS1,RS2} \  which
claims ``our world is confined to the 4-dimensional brane that is
embedded in higher dimensional space time''.  With respect to the
existence of higher dimensional space-time, this idea resembles
the old Kaluza-Klein compactification scenario,  but has great
difference from it that the wave function does not
spread over the extra dimension but has its support only on the
brane.  Various papers utilize such a $\delta$-distributional wave
function to explain the hierarchy or others.  Around such
a work,  one of the most surprising claim is that we do not need the
compactification to get the 4-dimensional gravity near the brane;
this is ``An
alternative to compactification'' by Randall and Sundrum\cite{RS2}
which we refer to as RS2 model henceforce.  So if you accept the
extra-dimension, then there are three ways to utilize it, an old
Kaluza-Klein, brane worlds with compact extra dimensions and with
non-compact ones.  But,  then,  ``What is the difference between
them?'',  or ``Cannot we see the difference except at the
high energy?''  To answer these questions,  we examine the Cauchy
problem of the gravitational wave with the initial data distributed
only around the brane in the RS2 model,  that is the third
possibility of the extra dimensions.  Our restriction to the initial
data is very natural in the brane world scenario since we are
confined on the brane and any physical phenomena takes place locally,
so if we disturb the space-time then such a signal is distributed only
around the brane.


The organization of this paper is as follows.  In the next section,
we review the RS2 model and formulate the problem which we want to
solve.  In the third section,  we solve and analyze the above Cauchy
problem.  Our main result is the $t^{-6}$ tail from the
$\delta$-distributional source of gravitational wave.  RS2 model is
also interesting from the fact that it has holographic dual
interpretation.\cite{G,DL,GKR,APR,RZ,P} \  We also make comments on
the above behavior from the holographic viewpoint.  In the final
section,  we give some comments and discussions.

\section{Setting the problem}
\label{sec:prob}

As discussed in the introduction,  we want to solve the Cauchy problem
of the gravitational wave in RS2 model.  So we first review RS2 and
set the problem appropriately.  We will mainly follow the framework
given by Ref. \citen{CR}.

In the RS2 model, we start with the five-dimensional manifold $M^5$
(and its mirror $\tilde{M}^5$)
with two-boundaries,  one is the brane boundary $\Sigma_1$ which is
introduced by hand and another is the infinite far-away boundary
$\Sigma_2$ from the brane.  So the Lagrangian is given by
\begin{align}
  \label{aln:lag}
  \inv{2\kappa_5^2} \left[\int_{M^5} d^5 x \sqrt{-g}(R + \Lambda_5)
    + \int_{\Sigma_1} d^4 x \sqrt{-\tilde{g}} {\cal L}_\textrm{matter}
    - 2 \int_{\p M^5} d^4 x \sqrt{-\tilde{g}} K \right].
\end{align}
We write the Lagrangian on the half of RS2 geometry.  Actually an
orientation reversed five-dimensional manifold $\tilde{M}^5$ must be
glued at the brane $\Sigma_1$.
In (\ref{aln:lag}),  $\kappa_5$ is the unit length of 5-dimensional
bulk,  $\Lambda_5$ is a 5-dimensional cosmological
constant,  $\tilde g$ is the induced metric on the boundary,  the
second term is
the matter Lagrangian and the third term represents the extrinsic
curvature contribution from the boundary $\p M^5=\Sigma_1\cup\Sigma_2$ 
called a Gibbons-Hawking term.\cite{GH} \ The necessity of the third
term is due to the requirement that,  in
the variational principle we only demand that the infinitesimal
variation itself vanishes uniformly along the infinite far-away
boundary, but the derivative of it does not vanish on the
boundary.\cite{Wa} \  Under such
a requirement the extrinsic curvature term is canceled with the
contribution of the derivative of variation from the bulk.  However in
the Lagrangian (\ref{aln:lag}),  we have also the brane boundary term
and we want non-zero fluctuation on the brane,  so above statement
does not hold on the brane side.  Under the appropriate 
infinitesimal variation,  the Lagrangian (\ref{aln:lag}) changes as,
\begin{align}
  \label{aln:var}
  \displaystyle
  \int_{M_5} d^5 x \sqrt{-g} \delta g^{\mu\nu} (R_{\mu\nu} -
  \f{g_{\mu\nu}}{2} R - \f{g_{\mu\nu}}{2}\Lambda_5)
  + \int_{\Sigma_1} d^4 x
  \sqrt{-\tilde{g}} \delta g^{\mu\nu} T_{\mu\nu}^\textrm{matt.} \nn \\
  - \int_{\Sigma_1} d^4 x \sqrt{-\tilde{g}} \delta
  g^{\mu\nu} \left(K_{\mu\nu}- \tilde{g}_{\mu\nu} K\right),
\end{align}
so that the first term leads to the equation of motion and the second
and third terms leads to the Israel junction condition\cite{I} due to
the fact $\delta g^{\mu\nu}\not=0$ on the brane:
\begin{align}
  \label{aln:Ein}
  G_{\mu\nu} - \inv{2} g_{\mu\nu} \Lambda_5 &= 0, \nn \\
  \Delta K_{\mu\nu} &= T_{\mu\nu}^\textrm{matt.} - \inv{3} g_{\mu\nu}
  T^\textrm{matt.}.
\end{align}
Here we use $\Delta K_{\mu\nu}$ as the difference of extrinsic
curvature across the brane.  The apearance of the differnce $\Delta
K_{\mu\nu}$ is due to the another contribution of the third term
in (\ref{aln:var}) from mirror manifold $\tilde{M}^5$.
This Israel junction condition with respect to $K_{\mu\nu}$ means that
the energy and the pressure on the
brane is balanced against the curvature of the bulk.  So when the
brane term consists only of the tension term ${\cal
  L}_\textrm{matter}=\lambda$,  we are able to get the RS2 geometry
which preserve the 4-dimensional Lorentz symmetry,
\begin{align}
  \label{aln:RSgeo}
  ds^2 = e^{(-2r_0-2\abs{r-r_0})/l} dx_4^2 + dr^2 , \nn \\
  -\f{e^{-2r_0/l}}{l} = \frac{\lambda}{12},~~~~~ \Lambda_5 = \f{12}{l^2},
\end{align}
if we require the $\Z_2$-orbifold condition
$K_{\mu\nu}(r=r_0+\epsilon)=-K_{\mu\nu}(r=r_0-\epsilon)$ .
Here we select the coordinate patch under which the brane is located
at $r=r_0$.

Then we consider the fluctuation around the above geometry.
If we vary the metric
$ds^2\rightarrow ds^2 + h_{\mu\nu}dx^\mu dx^\nu$ 
and require that this is also the solution of the equation of motion
with the boundary condition (\ref{aln:Ein}),  then we
get the equation of motion for the gravitational wave,
\begin{align}
  \label{aln:eom}
  \eta^{\mu\nu} \p_\mu \p_\nu \tilde h_{\rho\lambda}
  + \inv{l^2} \left(\p_{\tilde z}^2 - \f{3}{\tilde z}\p_{\tilde
      z}\right) \tilde h_{\rho\lambda} &= 0, \nn \\
  \p_{\tilde z} \tilde h_{\rho\lambda} &= 0 \qquad \textrm{at}~~
  \tilde{z} = \tilde{a}.
\end{align}
Here we consider the mode satisfying $h^\mu_{~\mu}=\p_\mu
h^\mu_{~\nu}=h_{5\mu}=h_{55}=0$ and have used the change of variables
$\exp\left(-2r/l\right)=1/{\tilde z}^2,
\exp\left(-2r_0/l\right)=1/{\tilde{a}}^2,
h_{\mu\nu}=\tilde h_{\mu\nu}/{\tilde z}^2$
in the region $r>r_0$.  In this expression,  we easily see that the
mode
$\tilde h_{\rho\lambda}(x^0,x^1,x^2,x^3)$ independent of
5-th coordinate $\tilde{z}$,
obeys the 4-dimensional massless equation of motion,  and in terms of
the original $h_{\mu\nu}$,  such a mode
is of the form $h_{\mu\nu}\propto\exp\left(-2r/l\right)$ so that the
gravity is ``localized'' around the brane.  However,  at this stage,
we wonder why such a definite form of gravitational wave can be
created by the four dimensional localized people.  In generic
situation,  when we create a disturbance just on the brane,  it is
physically natural to think that the created mode is not such a
harmonic mode on the 5-dimensional space-time but a localized
disturbance around the 4-dimensional brane\footnote{This is like the
  situation under which we live on the earth and make a disturbance on
  the ground in various ways.  Such a wave form of disturbance
  depends on how we create it,  the collision of a meteorite or the
  walking.  In comparison with such a situation,  the above context
  with localized gravity looks like a Rayleigh wave's one  a little.}.
So if we consider the interaction with the gravitation,  we must
consider not only the above harmonic stationary wave,  but also the
general initial condition distributed around the brane and solve its
asymptotic behavior to compare with
the 4-dimensional pure gravity.  So our problem is to solve
(\ref{aln:eom}) under the following initial condition,
\begin{align}
  \tilde h_{\mu\nu}(t=0,x,\tilde z) &= \tilde
  f_{\mu\nu}(x,\tilde z), \nn \\
  \p_t \tilde h_{\mu\nu}(t=0,x,\tilde z) &= \tilde
  k_{\mu\nu}(x,\tilde z).
\end{align}
where $\tilde{f}$ and $\tilde{k}$ has its support around the brane.
Of course,  in principle,  we must consider how the gravitational wave 
is created by the localized matter on the brane and then solve the
above equation,  but in this paper we don't consider such a problem
for simplicity.

\section{Analysis of asymptotic behavior}
\label{sec:asympt}

In the previous section,  we have formulated the Cauchy problem of
gravitational wave in RS2 model.  In this section,  we examine the
asymptotic behavior of the solution as $t\rightarrow\infty$.  For this
purpose,  we recall the way how to solve the Cauchy problem in general.
When we consider the Cauchy problem of some self-adjoint positive
definite operator $\Delta$,
\begin{align}
  \label{aln:hyperbolic}
  \f{\p^2}{\p t^2} \phi(x,t)  + \Delta \phi(x,t) &= 0, \\
  \intertext{with initial conditions}
  \phi(x,0) &= f(x), \nn \\
  \f{\p}{\p t} \phi(x,0) &= k(x),
\end{align}
we get the formal solution of it,
\begin{align}
  \label{aln:formsol}
  \phi(x,t) = \cos\sqrt{\Delta} t \, f(x) +
  \f{\sin{\sqrt{\Delta}}t}{\sqrt{\Delta}} k(x). 
\end{align}
However in this expression,  we need the spectral decomposition of
$\Delta$ to define $\sqrt{\Delta}$ explicitly.  This is 
not so difficult but the above expression is not useful to obtain the
asymptotic behavior of the solution,  so we want an another expression
of the solution.  This is done by the Duhamel principle.  The Duhamel
principle is stated as follows.  First we define $G(t) := \theta(t)
\sin\sqrt{\Delta t}/\sqrt{\Delta}$,  where $\theta(t)$ is the Heaviside
function.  Then,  taking care of the fact that the integration kernel
$G(t;x,y)$ of $G(t)$ at the region $t>0$ is the same as the solution
of (\ref{aln:hyperbolic}) with the initial condition 
$\phi(x,0)=0,\p_t\phi(x,0)=\delta(x-y)$,  we see that the
operator $G(t)$ is the solution of the equation,
\begin{align}
  \label{aln:Duhamel}
  \f{\p^2}{\p t^2} G(t)  + \Delta G(t) &= \delta(t)\mathbf{1}.
\end{align}
From this expression (\ref{aln:Duhamel}),  the form of $G(t)$ is given 
by Fourier-transforming $G(t)$ with respect to $t$,
\begin{align}
  \label{aln:funsol}
  \hat G(\omega) &= \inv{-(\omega+i0)^2 + \Delta}, \nn \\
  G(t) &= \int \f{d\omega}{2\pi} \hat{G}(\omega) e^{-i\omega t}.
\end{align}
Here $\omega+i0$ means that when we Fourier-transform $\hat G(\omega)$
back to $G(t)$,  we take the integration path so as to deform it into
the upper half plane at the spectrum of $\Delta$.  This is because the
operator $G(t)$ is only supported in the region $t>0$.  In
Ref. \citen{GKR}, 
a similar propagator as a solution of (\ref{aln:Duhamel}) is given,
but it is the Feynman propagator obtained by Wick rotation from
Euclidean space and does not suit our problem.  In
other words,  it obeys different boundary condition.  So,  we get
the operator $G(t)=\theta(t)\sin\sqrt{\Delta}t/\sqrt{\Delta}$,  then
we also get the operator $\theta(t) \cos\sqrt{\Delta}t$ by
differentiating $G(t)$ with respect to $t>0$.  Lastly we go back to
(\ref{aln:formsol}) and get the solution of (\ref{aln:hyperbolic})
with locally-distributed initial condition.
From the fact that the kernel $G(t;x,y)$ itself is almost the same as
the solution of (\ref{aln:hyperbolic}) with locally-distributed
initial conditions
$\phi(x,0)=0,\p_t\phi(x,0)=\delta(x-y)$,  we realize that $G(t;x,y)$
includes all the information of asymptotic behavior from the
localized disturbance.  So below we
analyze this kernel $G(t)$.  In addition to this fact,  we also
notice that when the localized source term is introduced,  then from
(\ref{aln:Duhamel}) we can see its contribution to the wave by simply
multiplying the kernel $G(t)$ on the source.  The Duhamel principle
works powerful independently how we create a wave.

Now let us return to our subject to examine the asymptotic behavior of
the gravitational wave.  We formulate our RS2 model so as to use the
Duhamel principle (\ref{aln:hyperbolic})-(\ref{aln:funsol}).
For such a purpose,  we transform $\tilde
h_{\mu\nu}$ to $\tilde{z}^{3/2}\phi$ and
$l\tilde{z}=z,l\tilde{a}=a$.  Then the equation of motion
(\ref{aln:eom}) with boundary condition changes into
\begin{align}
  \label{aln:hermite}
  \eta^{\mu\nu} \p_\mu \p_\nu \phi + (\p_z^2 -\f{15}{4 z^2}) \phi &=
  0, \nn \\
  \p_z \phi + \f{3}{2z} \phi &= 0 & \text{at}~~ z=a.
\end{align}
This change of variable $\tilde{h}_{\mu\nu}\rightarrow \phi$ is
necessary for achieving the
self-adjointness of $\Delta=-\delta^{ij}\p_i\p_j -\p_z^2+15/4z^2$ in
$L^2$-space.  Here $i,j$ run from one to three.  We choose $L^2$-space 
as the definition domain of $\Delta$ and this matches our
physical situation: The AdS space is
conformally flat and the speed of the signal is
finite.  Therefore,  when we create the disturbance around the brane,
such a signal does not reach at the boundary of the coordinate
system (\ref{aln:RSgeo}) in finite time.  So in the finite time
observation,  the signal is included in the $L^2$ space.  We also
ignore the spin dependence of $\tilde{h}_{\rho\lambda}$.  According to
the Duhamel principle,  we get the following retarded Green kernel
$\hat G(\omega)$ by the use of the eigenfunction expansion for this
differential equation system (\ref{aln:hermite}):
\begin{align}
  \label{aln:spcrep}
  & \hat G(\omega;\vec{x},z;\vec{y},w) \nn \\
  =& \int \f{d^3 \vec{k}}{(2\pi)^3 Z(a)}
  \f{e^{i \vec{k}\cdot (\vec{x}-\vec{y})} \phi_0(z)\phi_0(w)}
  {-(\omega+i0)^2 + \abs{\vec{k}}^2}
  + \int \sigma(\lambda) d\lambda \f{d^3 \vec{k}}{(2\pi)^3}
  \f{e^{i \vec{k}\cdot
      (\vec{x}-\vec{y})}\phi(\lambda,z)\phi(\lambda,w)}
  {-(\omega+i0)^2 + \abs{\vec{k}}^2 +\lambda}.
\end{align}
Here $\phi_0(z)$ and $\phi(\lambda,z)$ are the fifth dimensional
eigenfunctions and correspond to massless graviton and the
Kaluza-Klein mode respectively,  $\sigma(\lambda)$ is the fifth
dimensional spectral
measure with support $\left[0,\infty\right)$ depending on the
normalization of $\phi(\lambda,z)$,  and $Z(a)$ is a normalization
factor of $\phi_0(z)$:
\begin{align}
  \label{aln:spec}
  \sigma(\lambda) &= \f{2}{\pi^2} \inv{\lambda a}
  \inv{Y_1^2(\sqrt{\lambda}a) + J_1^2(\sqrt{\lambda}a)}, \qquad
  Z(a) = \f{a}{2},
  \nn \\
  \phi_0(z) &= \left(\f{a}{z}\right)^\f{3}{2},
  \nn \\
  \phi(\lambda,z) &= \f{\pi}{2} \sqrt{\lambda az}
  \left(Y_1(\sqrt{\lambda}a) J_2(\sqrt{\lambda}z) -
    J_1(\sqrt{\lambda}a) Y_2(\sqrt{\lambda}z) \right),
\end{align}
and $J_\mu(z),Y_\mu(z)$ are the Bessel functions.  The derivation of
this eigenfunction expansion (\ref{aln:spcrep}) is given 
in Appendix \ref{sec:delta}.

So we can get the analytic expression of the solution of Cauchy
problem,
then we estimate the asymptotic behavior $t\rightarrow\infty$ of this 
expression on the brane $z=w=a$.  The first term of (\ref{aln:spcrep})
has a well known analytic formula so here we estimate the second term
$\hat{G}_\textrm{KK}$.  At the brane $z=a$,  the eigenfunctions
$\phi(z;\lambda)$ become a constant independent of $\lambda$ (really
these are $1$) so we
can concentrate on the spectral measure $\sigma(\lambda)$ as
calculated below.  In the following,  we can set $\vec{y}=0$ so that
the gravitational wave is created at the origin on the brane for
convenience.

\begin{figure}
  \centerline{\includegraphics[height=4cm]{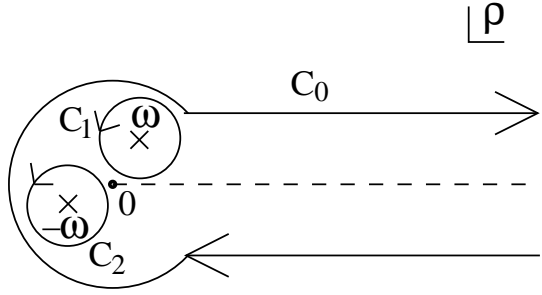}}
    \caption{The contribution to the singular support from the
      spectral of $\Delta$.  Here $\omega$ really means $\omega+i0$
      which is a pole of $F(\rho)$.}
    \label{fig:non}
\end{figure}

Our main technique is to use the Paley-Winner-Schwartz theorem
which claims that the Fourier transformation of a compactly supported
infinitely differentiable function decreases faster than any power, 
so that the only contribution which is responsible for the asymptotic
behavior comes around the singular support\footnote{A singular
  support is defined as the closure of the points at which the
  function is not infinitely differentiable.}  of the original
function.  Back to the second term of (\ref{aln:spcrep}),  and we want
to know the asymptotic behavior of the Fourier transformation of it,
so we must look for the singular support with respect to $\omega$.
This is easily found,  and $\omega=0$ is the only candidate point
which belongs to the singular support of
$\hat{G}_\textrm{KK}(\omega)$.  Other point is holomorphic with
respect to $\omega$.  We also notice that when we approach to
$\omega=0$ along,  for an example,  the pure imaginary axis,  then the
$\int d^3k d\lambda$-integration around the origin
$\rho^2:=k^2+\lambda=0$ is convergent by power counting so that the
point $\omega=0$ is not the pole but the branch point of a function
$\hat{G}_\textrm{KK}(\omega)$.  So
we analyze what branch appear after the $\int d^3k
d\lambda$-integration.
This is done as follows.  First we change the integrand $f(\rho)$ of
the second term of (\ref{aln:spcrep}) (here we suppress the
integration variables except for $\rho$) into an analytic function
$F(\rho)$ on $\C\setminus\R_+$ such that discontinuity over the
positive real axis is $f(\rho)=F(\rho+i0)-F(\rho-i0)$,  and also
change the $\rho$-integration path into the clockwise curve around the
real axis.  Then we can further deform the $\rho$-integration path to
$C_0+C_1+C_2$ of figure \ref{fig:non}.
For the integration along the path $C_0$,  $\omega\rightarrow 0$ limit
is no longer
singular,  and the singularity of the original $\int d^3k
d\lambda$-integration concentrates on the $\rho$-integration along the
small closed curve $C_1$ and
$C_2$.  This last integration is easily performed and if we write the
integrand in the form
$F(\rho)=\Sigma(\rho)/\left(-(\omega+i0)^2+\rho^2\right)$,
then the contribution from the paths $C_1$ and $C_2$ is
\begin{align}
  \label{aln:con}
  \f{\pi i}{\omega} \left(\Sigma(\omega)-\Sigma(-\omega)\right).
\end{align}
Because we search singular support,  the holomorphic part of
(\ref{aln:con}) with respect to $\omega$ can be ignored.
In this way,  after performing the integration of remaining variables,
we get the singularity of $\hat{G}_\textrm{KK}(\omega)$ as
\begin{align}
  \label{aln:GKK}
  \hat{G}_\textrm{KK}(\omega) &\rightarrow i \f{4}{5!\pi} a^3
  \omega^5 \log\omega + \cdots,
\end{align}
in the small $\omega$ limit.  The actual calculus is performed in
Appendix \ref{sec:singular}.  Here $\omega$ really means $\omega+i0$.
We then come back to the time
representation of $G_\textrm{KK}(t)$ and get the asymptotic
behavior
\begin{align}
  G_\textrm{KK}(t,\vec{0},a;\vec{0},a) \rightarrow -\f{4}{\pi}
  \f{a^3}{t^6} \theta(t) + \cdots.
\end{align}
If we take account of the Lorentz
covariance on the brane,  we finally find the Kaluza-Klein
contribution to asymptotic behavior
\begin{align}
  G_\textrm{KK}(t)
  &\rightarrow -\f{4}{\pi} \f{a^3}{(t^2-\abs{\vec{x}}^2)_+^3} +
  \cdots,
\end{align}
in the large $t$ limit.  Here the suffix $+$ means that the support is
only on the $t>\abs{\vec{x}}$ by the causality.  This causality could
be checked by the $\abs{\vec{x}}$ dependent term in (\ref{aln:GKK})
which is higher order in $\omega$. 

In summary,
including also the contribution from the massless graviton mode,
the leading asymptotic behavior from localized disturbance is
\begin{align}
  \f{\delta(t-\abs{\vec{x}})}{2\pi a \abs{\vec{x}}}
  + \cdots
  - \f{4}{\pi} \f{a^3}{(t^2-\abs{\vec{x}}^2)_+^3} + \cdots .
\end{align}
The first term is an usual four-dimensional retarded Green function
which comes from the graviton and represents the Huygens principle in
4-dimensional space-time,
that is to say the wave form is determined only by its past
light-cone.  The second term is our main result which comes from the
continuous spectrum of Kaluza-Klein modes and represents the breaking
of
Huygens principle,  that is to say the wave form depends on all of
the causal past region.  This breaking of Huygens principle which arose 
from the existence of extra dimension,  can be
compared with the flat 5-dimensional space-time's one without brane.
In the usual flat 5-dimension,  the retarded Green function behaves as
$t^{-3}$.  The difference from our case shows the property that in RS2
model,  the Kaluza-Klein modes highly decouple from the brane mode as
noted in Ref. \citen{RS2}.  But when we compare this effect with the
compact brane world case,  then in compact case the
contribution from the \emph{discrete} Kaluza-Klein modes lead to
asymptotically exponential decay.  So the power decay behavior of
gravitational  wave distinguishes the noncompact brane world from
compact one's.  The  difference between them is caused by whether or
not there exists continuous spectrum as seen by our calculus.  This
continuous spectrum, so the power decay of the wave, is a
model independent feature of non-compact brane world which 
generically holds even
if the continuous spectrum begins after some gap (in such a case the
simple power decay is replaced by power decay with oscillation.).
Also this power  decay feature is not restricted to the gravitational
wave,  but extended to any other localized mode which comes from the
field in the non-compact bulk.  So even at the classical level and low
energy,  we can distinguish non-compact brane world from compact one
(if the power decay behavior different from pure 4-dimensional's one
is observed).  Also we can conclude that presently observed fields
classically like electro-magnetic field, must be a non-bulk field
living on the brane or must propagate on small compactified space as
far as they satisfy the Huygens principle in the category of power
decay behavior.  In the old Kaluza-Klein
cases without brane,  local disturbance in such
circumstance cannot be created so that the power decay behavior in our
calculus distinguishes itself from them.

So far,  we have considered the asymptotic behavior of the
gravitational wave which is initially distributed around the brane,
but then it is interesting to ask whether or not such a signal can be
observed in a real experiment.  As is claimed before, 
more realistic signal depends on how it is created by,  for an
example,  a binary star.  We do not pursue such a thing,  but we
estimate the modification of Newton law by the same mode as the above
$t^{-6}$ contribution.  In the process of deriving
the asymptotic behavior,  the essential thing is that the equation
(\ref{aln:spcrep}) with $z=w=a$ is like the K\"all\'en-Lehmann
spectral representation except for the prescription of the integration 
path,
\begin{align}
  \label{aln:KL}
  \hat{G}(\omega,x,a;y,a) &= \int \f{d^3 k}{(2\pi)^3}
  \f{Z(a)^{-1}e^{i\vec{k}\cdot(\vec{x}-\vec{y})}}{-(\omega+i0)^2 + k^2}
  + \int \f{d^3 k}{(2\pi)^3} d\mu
  \f{\tilde{\sigma}(\mu^2)e^{i\vec{k}\cdot(\vec{x}-\vec{y})}}
  {-(\omega+i0)^2 + k^2+\mu^2},
  \nn \\
  \tilde{\sigma}(\mu^2) d\mu &=
  \f{4\mu d\mu}{\pi^2 a \mu^2\left(N_1^2(\mu a) + J_1^2(\mu a)\right)},
\end{align}
and the spectral function behaves in the small $\mu$ limit as
\begin{align}
  \label{aln:spasy}
  \tilde{\sigma}(\mu^2) &\rightarrow a\mu + \f{a^3\mu^3}{2}(2\gamma-1)
  +a^3\mu^3\log\f{a\mu}{2} + \cdots
\end{align}
Our calculus is concerned with the third term of (\ref{aln:spasy}),
the first and second term does not contribute to asymptotic behavior
of gravitational wave as seen in Appendix \ref{sec:singular}.  When we
compare with the Newton law,  the first term mainly contributes and
leads to $1/r^3$ potential as is calculated in Ref. \citen{RS2}.  Our
third term has tinier effect $\log
r/r^5$ to the gravitational potential,  so unfortunately it is difficult
to observe the signal of RS2 model by the present gravitational
experiment.

Finally,  we comment on the complementarity between AdS/CFT
correspondence\cite{M,GKP,W}(see a review \citen{MAGOO} for an example)
and RS2 geometry.  Its original motivation\cite{G} is the
above $1/r^3$ potential,  which is equivalent to the
graviton-conformal mode-graviton diagram.  Here conformal mode means
two point function
of stress tensor of dual conformal field theory.  This interpretation
is explicit in the Feynman prescription version of
(\ref{aln:KL}) and its expansion (\ref{aln:spasy}).  As far as we are
concerned with the four-momentum dependence,  the first term of
(\ref{aln:spasy}) corresponds to the above 
mentioned diagram,  and the second term represents graviton to
conformal mode diagram and the third term which is responsible for the
asymptotic behavior of gravitational wave,  represents graviton to
conform mode to graviton to conformal mode to graviton diagram (this is
$k^2\log k$ expansion.).  So the complementarity between AdS/CFT
correspondence and RS2 geometry does work qualitatively in our
calculus.

\section{Comments and Discussions}
\label{sec:end}

In this paper,  we have considered the Cauchy problem in the RS2 brane
world and analyzed the asymptotic behavior of the gravitaional wave
created by the localized source.  The power decay behavior resulting
from the continuous Kaluza-Klein mode,  which is universal
in noncompact brane world represents the
difference (although being difficult to observe it now) even at the
classical level from a compact brane world,  in
which case the gravitational wave decays exponentially by a mass gap.
It also gives a distinction from a Kaluza-Klein scenario,  in which
case we are uniformly
distributed over extra dimensions so that no local disturbance cannot
be created.
We also find that in a real calculus,  we only need the behavior of
spectral function of Kaluza-Klein modes around zero or a mass gap.
This behavior leads to the correction
of Newton law or the modification to asymptotic behavior of
gravitational wave.  Especially,  in RS2 model,  we can get the
meaning of spectral function from AdS/CFT correspondence.  However at
this point,  we have a natural question.  When we are confined in
the brane,  how can we get the information like curvature or the shape
of extra dimension?  Such an information is the most important one so
that we determine the spectral function.  Against this question,  we
have a partial answer in one extra-dimensional case.  If the spectral
function is determined,  then
we get the potential  (in our case,  the curvature of extra
dimensions) by solving the Gel'fand-Levitan equation.\cite{GL} \  So
the remaining question is how we observe the full spectral function
not restricted to the information around zero.  This is a typical
inverse problem and if we discover the extra dimension in future,  it
will be important to
determine the curvature of extra dimension from the signal.  Of
course,  from the point of view of inverse problem,  more
practical method is desired.  For example,  high frequency expansion
(eikonal expansion) of gravitational wave $\tilde{h}_{\mu\nu}\propto
e^{i\omega S(t,x,z)} \sum_j \omega^{-j} a_j(t,x,z)$ over any
five-dimensional metric $ds^2=e^{2A(z)}(dx_4^2+dz^2)$ tells us that the
leakage rate per unit proper time is proportional to $m
\sqrt{G_{zz}}$,  here $m$ means that the Kaluza-Klein mass which leaks
to fifth dimension and $G_{zz}$ means the fifth Einstein tensor around
the brane.  Among such things or combination of them,  what methods
are more appropriate to explore the extra dimension?  And we ask
``Can one feel a shape of extra dimension?''.


\section*{Acknowledgments}
I would like to thank T. Kugo for careful reading of the manuscript.

\appendix
\section{Eigenfunction expansion of the operator in
  (\ref{aln:hermite})}
\label{sec:delta}

In this appendix, we will explain the (generalized) eigenfunction
expansion of the $z$-part of the differential operator in
(\ref{aln:hermite}) and its completeness .  Generalizing the
operator a little bit,  we here consider the self-adjoint differential
operator,
\begin{align}
  \label{aln:op}
  P(x,D) &= -\f{d^2}{dx^2} + \f{\nu^2-1/4}{x^2},
\end{align}
with the boundary condition,
\begin{align}
  \label{aln:opbc}
  \psi'(a) = \f{1/2-\nu}{a} \psi(a).
\end{align}
We work in $L^2\left(\left[a,\infty\right)\right)$ space with boundary 
condition (\ref{aln:opbc}), and Eq. (\ref{aln:hermite}) corresponds to
the $\nu=2$ case.  Our treatment is
standard (see \citen{RYT} for an example) and the outline is as
follows.  First
we construct the Green function $G(x,y;\lambda)$ which is the
integration kernel of the
resolvent $G(P;\lambda)=(\lambda -P(x,D))^{-1}$.  Here $\lambda\in\C$
is included in the complement of the spectrum of $P(x,D)$.  This Green
function is given by the solutions of
differential equation $P(x,D)\psi(x)=\lambda\psi(x)$ as explicitly
calculated below.  Then using the delta function formula,
\begin{align}
  \label{aln:deltadef}
  \delta(x) = - \lim_{\epsilon\rightarrow 0} \inv{2\pi i}
  \left(\inv{x+ i\epsilon} - \inv{x - i\epsilon} \right) ,
\end{align}
we get the desired form of eigenfunction expansion (the spectral
decomposition),
\begin{align}
  \label{aln:opeigen}
  f(x) &= \int_\R d\lambda \ \delta(\lambda - P(x,D)) f(x) \nn \\
  &= -\int_\R d\lambda \lim_{\epsilon\rightarrow 0} \inv{2\pi i} \left[
    G(P;\lambda+i\epsilon) - G(P;\lambda-i\epsilon)\right] f(x) .
\end{align}
Below we will write this fact explicitly in the form
\begin{align}
  \delta(x-y)=-\frac{1}{2\pi i} \lim_{\epsilon\rightarrow 0} \int_\R
  d\lambda\left[G(x,y;\lambda+i\epsilon)-G(x,y;\lambda-i\epsilon)\right]
\end{align}

In order to construct the Green function,  we consider the
differential equation,
\begin{align}
  \label{aln:diff}
  P(x,D) \psi(x) &= \lambda \psi(x) , \qquad \lambda \in \C,
\end{align}
Take two independent solutions of (\ref{aln:diff}) with following
initial conditions,
\begin{align}
  \label{aln:bc}
  \psi_1(a) = \cos\alpha, && \psi_1'(a) = \sin\alpha, \nn \\
  \psi_2(a) = -\sin\alpha, && \psi_2'(a) = \cos\alpha .
\end{align}
Here $\alpha$ is arbitrary in principle  and in the case of
(\ref{aln:opbc}),  we take,
\begin{align}
  \cos \alpha = \f{a}{\sqrt{a^2+ (1/2-\nu)^2}} , &&\dis
  \sin \alpha = \f{1/2-\nu}{\sqrt{a^2+ (1/2-\nu)^2}} .
\end{align}
Then $\psi_1$ is the solution of (\ref{aln:diff}) with boundary
condition (\ref{aln:opbc}).  In our case the solutions $\psi_1$ and
$\psi_2$ are given explicitly by Bessel functions,
\begin{align}
  \psi_1(x;\lambda) &= \sqrt{x}\left[A_\lambda J_\nu(\sqrt{\lambda}x)
    + B_\lambda Y_\nu(\sqrt{\lambda}x)\right], \nn \\
  \psi_2(x;\lambda) &= \sqrt{x}\left[C_\lambda J_\nu(\sqrt{\lambda}x)
    + D_\lambda Y_\nu(\sqrt{\lambda}x)\right].
\end{align}
Here $A_\lambda,B_\lambda,C_\lambda,D_\lambda$ are constants,
\begin{align}
  A_\lambda &=
  \f{\pi a^\inv{2}}{2} \left[Y_\nu(\sqrt{\lambda} a)
    \left(-\sin\alpha+\inv{a}(\inv{2}-\nu)\cos\alpha\right)
    +\sqrt{\lambda}Y_{\nu-1}(\sqrt{\lambda} a)\cos\alpha\right], \nn \\
  B_\lambda &= 
  \f{\pi a^\inv{2}}{2} \left[J_\nu(\sqrt{\lambda} a)
    \left(\sin\alpha-\inv{a}(\inv{2}-\nu)\cos\alpha\right)
    -\sqrt{\lambda}J_{\nu-1}(\sqrt{\lambda} a)\cos\alpha\right], \nn \\
  C_\lambda &= 
  \f{\pi a^\inv{2}}{2} \left[Y_\nu(\sqrt{\lambda} a)
    \left(-\cos\alpha-\inv{a}(\inv{2}-\nu)\sin\alpha\right)
    -\sqrt{\lambda}Y_{\nu-1}(\sqrt{\lambda} a)\sin\alpha\right], \nn \\
  D_\lambda &=
  \f{\pi a^\inv{2}}{2} \left[J_\nu(\sqrt{\lambda} a)
    \left(\cos\alpha+\inv{a}(\inv{2}-\nu)\sin\alpha\right)
    +\sqrt{\lambda}J_{\nu-1}(\sqrt{\lambda} a)\sin\alpha\right].
\end{align}

We also need the fast-decreasing function at infinity for
$\lambda\in\C\setminus\R$ ,  because we consider the $L^2$ space.
We write such a function
$\psi^{(1)}(x;\lambda)$ for $\Im\lambda>0$,  and
$\psi^{(2)}(x;\lambda)$ for $\Im\lambda<0$.  They are given by
\begin{align}
  \psi^{(1)}(x;\lambda) &= \sqrt{x}\left[J_\nu(\sqrt{\lambda}x)
  + i Y_\nu(\sqrt{\lambda}x)\right], \nn \\
  \psi^{(2)}(x;\lambda) &= \sqrt{x}\left[J_\nu(\sqrt{\lambda}x)
  - i Y_\nu(\sqrt{\lambda}x)\right].
\end{align}

Using these,  we can construct the Green function
$G^\pm(x,y;\lambda)$,
\begin{align}
  \label{aln:Green}
  G^+(x,y;\lambda) &= \inv{W[\psi_1,\psi^{(1)}]}
  \left\{
    \begin{array}[h]{cc}
      \psi_1(x) \psi^{(1)}(y) & x<y\\
      \psi^{(1)}(x) \psi_1(y) & x>y
    \end{array}
  \right. \nn \\
  &= 
  \left\{
    \begin{array}[h]{cc}
      \dis
      -\f{C_\lambda+i D_\lambda}{A_\lambda+i B_\lambda}
      \psi_1(x)\psi_1(y) + \psi_1(x) \psi_2(y) & x<y \\
      \dis
      -\f{C_\lambda+i D_\lambda}{A_\lambda+i B_\lambda}
      \psi_1(x)\psi_1(y) + \psi_2(x) \psi_1(y)& x>y
    \end{array}
  \right. \qquad \textrm{for} ~\Im\lambda>0,
  \nn \\
  G^-(x,y;\lambda) &= \inv{W[\psi_1,\psi^{(2)}]}
    \left\{
      \begin{array}[h]{cc}
        \psi_1(x)\psi^{(2)}(y) & x<y\\
        \psi^{(2)}(x)\psi_1(y) & x>y
      \end{array}
    \right. \nn \\
    &= \left\{
      \begin{array}[h]{cc}
        \dis
        -\f{C_\lambda-i D_\lambda}{A_\lambda-i B_\lambda}
        \psi_1(x)\psi_1(y) + \psi_1(x) \psi_2(y) & x<y \\
        \dis
        -\f{C_\lambda-i D_\lambda}{A_\lambda-i B_\lambda}
        \psi_1(x)\psi_1(y) + \psi_2(x) \psi_1(y)& x>y
      \end{array}
    \right.\qquad \textrm{for} ~\Im\lambda<0 .
\end{align}
Here $W[\psi,\phi]=\psi(x)\phi'(x)-\psi'(x)\phi(x)$ is a Wronskian,
so it is a constant and explicitly gives
$W[\psi_1,\psi^{(1)}]=2i(A_\lambda+iB_\lambda)/\pi$ and
$W[\psi_1,\psi^{(2)}]=-2i(A_\lambda-iB_\lambda)/\pi$.
It is easy to check that this Green function $G^\pm(x,y;\lambda)$
certainly maps the $L^2$ space with boundary condition
(\ref{aln:opbc}) to itself by the fact that $\psi_1$ obeys the boundary
condition and $\psi^{(1)},\psi^{(2)}$ are fast-decreasing functions for
$\pm\Im\lambda>0$ respectively.

So we get the Green function,  then consider the eigenfunction
expansion.  In such a construction,  an essential point is that the
solution of (\ref{aln:diff}) with $\lambda$-independent
boundary conditions like (\ref{aln:bc}) has the holomorphic
dependence on $\lambda\in\C$.  This statement is easily checked in
our case and also proved in more general setting.  And so,  to
construct the eigenfunction expansion using the formula
(\ref{aln:opeigen}),
it is convenient to use the solution $\psi_1,\psi_2$, not
$\psi^{(1)},\psi^{(2)}$, that is the second expression of $G^\pm$ in
(\ref{aln:Green}).
Then we can easily see the pole (which corresponds to the discrete
spectrum) and the discontinuity across the real axis (which corresponds
to the continuous spectrum) of the function $G^+ - G^-$ with respect
to $\lambda\in\R$ because such an information concentrates on the
coefficient of
$\psi_1(x)\psi_1(y)$ of (\ref{aln:Green}).  Finally we get the delta
function in the case of $-\sin\alpha+\inv{a}(\inv{2}-\nu)\cos\alpha
\ne 0$,
\begin{align}
  \delta(x-y) &=
  \inv{2} \int_0^\infty d\lambda \inv{A_\lambda^2+B_\lambda^2}
  \psi_1(x;\lambda)\psi_1(y;\lambda).
\end{align}
In the case of $-\sin\alpha+\inv{a}(\inv{2}-\nu)\cos\alpha = 0$,  we
get an extra pole term and this leads to
\begin{align}
  \delta(x-y) &= \inv{Z(a)} \psi_0(x)\psi_0(y)
  + \inv{2} \int_0^\infty d\lambda \inv{A_\lambda^2+B_\lambda^2}
  \psi_1(x;\lambda)\psi_1(y;\lambda).
\end{align}
Here $Z(a)$ is a normalization factor and $\psi_0(x)$ is a
$\lambda\rightarrow 0$ limit of $\psi_1(x;\lambda)$.  Their explicit
forms are given by
\begin{align}
  Z^{-1}(a) &= 
  \f{2}{a} \left(1+\inv{a^2}(\inv{2}-\nu)^2\right)(\nu-1), \nn \\
  \psi_0(x) &= \left(\f{a}{x}\right)^{\nu-\inv{2}} \cos\alpha.
\end{align}
This is the desired form in (\ref{aln:spcrep}) in the text.  The
functions $\phi_0(z)$ and $\phi(\lambda,z)$ used there are
$\psi_0(z)/\cos\alpha$ and $\psi_1(z;\lambda)/\cos\alpha$
respectively in the case of $\nu=2$.

\section{The calculus of singularity of Green kernel
  (\ref{aln:spcrep})}
\label{sec:singular}

In this appendix,  we will give the derivation of
(\ref{aln:GKK}) which is an expression of singularity of
(\ref{aln:spcrep}).  First  we change the integration variables in
(\ref{aln:spcrep}) to
\begin{align}
  \sqrt{\lambda} &= \rho \cos\theta, & \vec{k} &= \rho \sin\theta
  \vec{\Omega_2}.
\end{align}
The range of variables are
$\rho\in\left[0,\infty\right),\vec{\Omega}_2\in \textrm{S}^2$ and
$\theta\in\left[0,\pi/2\right]$.  Then the the
second term of (\ref{aln:spcrep}) becomes
\begin{align}
  \label{aln:integrand}
  \int \f{\sigma(\sqrt{\rho\cos\theta})
    e^{i\rho\sin\theta\vec{\Omega}_2\cdot\vec{x}}}{
    -(\omega+i0)^2+\rho^2}
  \f{\rho^4\cos\theta \sin^3\theta d\rho d\theta d\Omega_2}{4\pi^3}.
\end{align}
Here $d\Omega_2$ is the volume form of 2-sphere.  We want to express
the integrand of (\ref{aln:integrand}) as the discontinuity of
$F(\rho)=\Sigma(\rho)/\left(-(\omega+i0)^2+\rho^2\right)$ with respect 
to $\rho$.  However the full form of $F(\rho)$ and $\Sigma(\rho)$ is
not required:  As explained in the
text,  the singularity of (\ref{aln:integrand}) is concentrated on the
expression (\ref{aln:con}) and we also know that the higher
power of $\omega$ is Fourier-tansformed to less power of $t$,  so 
to get asymptotic behavior with respect to $t$ (it is original
motivation in the text),  it is enough to examine the
first few singular terms in (\ref{aln:con}),  in other words the first
few terms 
of $\Sigma(\rho)$.  By definition,  the first few terms of
$\Sigma(\rho)$ come from the ones of the numerator of integrand in
(\ref{aln:integrand}).  Really the numerator behaves in the small
$\rho$ limit as
\begin{align}
  \label{aln:speasy}
  \left(\f{a}{2}  +
    \f{a^3(2\gamma-1)}{4} \rho^2 \cos^2\theta
    + \f{a^3}{2} \rho^2 \cos^2 \theta \log \right. & \left. 
    \f{a\rho\cos\theta}{2}
    +  \cdots \right) \nn \\
  & \times e^{i\rho\sin\theta\vec{\Omega}_2\cdot\vec{x}}
  \f{\rho^4 \cos\theta \sin^2\theta d\theta d\Omega_2}{4\pi^3}.
\end{align}
This expansion corresponds to the one of (\ref{aln:spasy}).  From
(\ref{aln:speasy}) we easily read the form of $\Sigma(\rho)$.  However
we only need the singularity of (\ref{aln:con}) so again the full
expression is not required,  for the even function part of
(\ref{aln:speasy}) does not contribute to the singularity of
(\ref{aln:con}).  After all,  the following expression which
discontinuity gives the third term in parenthesis of
(\ref{aln:speasy}),  is needed instead of $\Sigma(\rho)$:
\begin{align}
  \label{aln:esigma}
  -\f{a^3}{8\pi i} \rho^2 \cos^2\theta (\log\rho)^2
  + \f{a^3}{4} \rho^2 \cos^2\theta \log\rho.
\end{align}
Here we take the cut of logarithm along the positive real axis and we
ignore the function which discontinuity gives the even function part
of third term in (\ref{aln:speasy}) by the same reason above.  We make 
comments on above abandonment of the even function part.  In the
flat 4-dimensional case,  the same procedure as above only gives
the even function part,  so no power decay behavior exists and it is
consistent with the Huygens principle.  In the flat five dimensional
case,  the odd function appears so that there exists power decay
behavior $t^{-3}$ as noted down in the text.  Then we go back to the
singularity of (\ref{aln:con}).  It is given by substituting
(\ref{aln:esigma})  into (\ref{aln:con}) with measure in
(\ref{aln:speasy}):
\begin{align}
  \int \f{i a^3}{16\pi^2} \omega^5 \log\omega \ 
  e^{i\omega \sin\theta \vec{\Omega}_2 \cdot \vec{x}}
  & \cos^3\theta \sin^2\theta d\theta d\Omega_2 \nn \\
  &= \int_0^{\frac{\pi}{2}} \f{i a^3}{4\pi}\omega^5 \log\omega
  \f{\sin(\omega\abs{\vec{x}}\sin\theta)}{\omega\abs{\vec{x}}\sin\theta}
  \cos^3\theta \sin^2\theta d\theta.
\end{align}
This gives (\ref{aln:GKK}) in the small $\omega$ limit as desired.


%

\end{document}